\documentclass[twocolumn,groupedaddress]{revtex4}  

\usepackage{graphicx, color}  
\usepackage{fancyhdr}
\usepackage{dcolumn}   
\usepackage{bm}        
\usepackage{amsmath,amssymb,siunitx}
\usepackage[version=3]{mhchem}
\usepackage{upgreek}
\usepackage{lastpage,siunitx}
\definecolor{linkcolor}{rgb}{0,0,0.6} 

\usepackage{xr}

\begin{document}

\title{Dynamic optical rectification and delivery of active particles}

\author{N. Koumakis}
\author{A. T. Brown}
\author{J. Arlt}
\author{S. E. Griffiths}
\author{V. A. Martinez}
\author{W. C. K. Poon}
\affiliation{SUPA and School of Physics \& Astronomy, The University of Edinburgh, James Clerk Maxwell Building, Peter Guthrie Tait Road, Edinburgh EH9 3FD, Scotland, UK}

\begin{abstract}
We use moving light patterns to control the motion of {\it Escherichia coli} bacteria whose motility is photo-activated. Varying the pattern speed controls the magnitude and direction of the bacterial flux, and therefore the accumulation of cells in up- and down-stream reservoirs. We validate our results with two-dimensional simulations and a 1-dimensional analytic model, and use these to explore parameter space. We find that cell accumulation is controlled by a competition between directed flux and undirected, stochastic transport. We articulate design principles for using moving light patterns and light-activated micro-swimmers to achieve particular experimental goals.
\end{abstract}

\maketitle

Active colloids~\cite{FermiColloids} are fundamentally interesting, exhibiting phenomena not found in equilibrium systems such as currents and pattern formation~\cite{CatesReview2012}. These have been applied, e.g., to concentrate particles~\cite{Galajda2007} or separate them by size, actuate micro-machines~\cite{di2010bacterial}, or self-assemble microstructures. For example, V-shaped `funnel gates' fabricated using soft lithography can rectify the motion of randomly-swimming bacteria \cite{Galajda2007}, producing steady currents or spatial patterns. The same effect can be achieved by applying a spatial light pattern to bacteria or other active colloids whose speed $v$ depends on the intensity of incident light, $I$. If ${\mathrm d}v/{\mathrm d}I >0$, the cell density $\rho$ builds up in darker regions, because $\rho v =$~constant \cite{TailleurPRL}.

This technique for `painting patterns with bacteria' has been demonstrated using {\it Escherichia coli} in which the proton motive force (PMF) driving swimming is generated by light-powered proteorhodopsin (PR) \cite{Jochen2018,Frangipane2018}. Potential applications include directing swimmers into compartments \cite{Galajda2007,Koumakis2013} and actuating micro-mechanical components \cite{di2010bacterial}. Such `bacterial painting' becomes significantly more versatile if the template is dynamic. Thus, globally time-varying light fields projected onto PR-driven {\it E. coli} can `erase' and `re-paint' patterns \cite{Jochen2018}. Here, we study the response of PR-driven {\it E. coli} to spatio-temporally varying light fields, specifically, wave-like propagating periodic patterns of illumination. By using a combination of experiments, theory and simulations we uncover a rich array of often counter-intuitive phenomena. 

\begin{figure}[t]
\includegraphics[width=0.5\textwidth,clip]{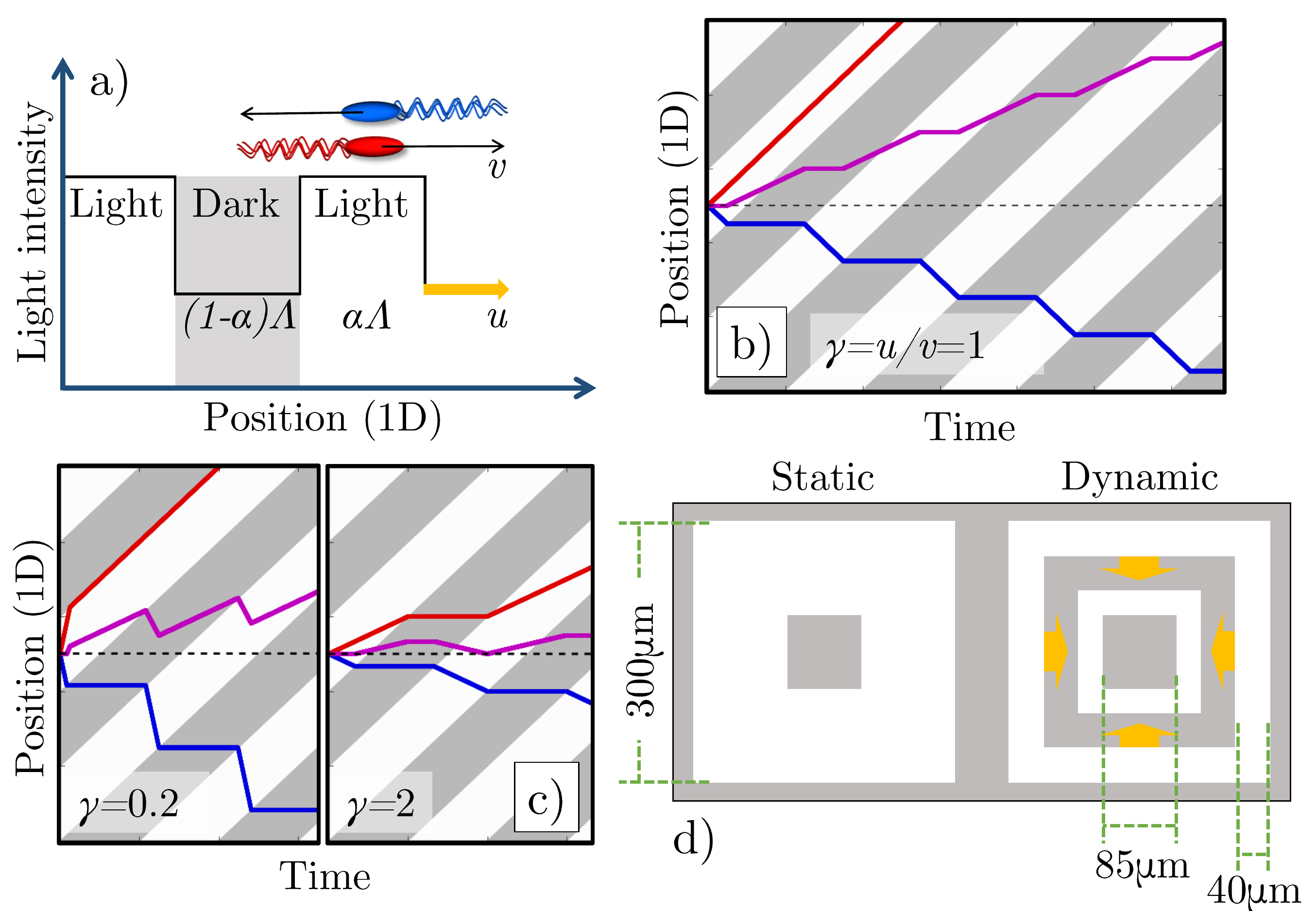}
\caption{\label{fig:Fig1} (a) 1D schematic of a translating square-wave light field with co- (red) and counter-moving (blue) bacteria. (b) Kymograph of a light field moving at speed $u$ (duty cycle $\alpha=0.5$) with light regions as white and dark ones with grey stripes and with trajectories of non-tumbling right (red) and left (blue) swimmers at speed $v= u$ (i.e.~$\gamma\equiv u/v= 1$) and the mean trajectory (magenta). (c) Kymograph for $\gamma=0.2$ and 2, and $\alpha = 0.5$. (d) Schematic of the 2D light patterns in our experiments.} 
\end{figure}  

To motivate our work, consider a 1D model solvable by inspection. Bacteria swim right ($+x$, $\rightharpoonup$) or left ($-x$, $\leftharpoondown$) at speed $v$ when illuminated, and stop completely in the dark; the (equal) $\rightharpoonup$ and $\leftharpoondown$ populations do not exchange. Now impose and translate at speed $u> 0$ a square-wave light pattern, Fig.~\ref{fig:Fig1}a. If $\gamma = u/v = 1$, $\rightharpoonup$ cells keep up with the moving light field in steady state, and so maintain their speed $v$, whereas $\leftharpoondown$ cells spend only some of their time in the light, determined by the duty cycle (fractional on time) of the pattern, $\alpha$. This results in a net positive cell flux, i.e.~a light pattern translating at $\gamma = 1$ rectifies the cells' motion and transports them with the pattern. We can represent this behaviour in a space-time plot (kymograph), Fig.~\ref{fig:Fig1}b. Optimal rectification is obtained for $\gamma \approx 1$: for $\gamma\ll 1$, $\rightharpoonup$ cells are trapped at the light-dark interface, whereas for $\gamma\gg 1$, both $\rightharpoonup$ and $\leftharpoondown$ cells spend $\approx 50\%$ time in the dark and the light. Either way, the net flux falls, Fig.~\ref{fig:Fig1}c.

Our experiments reproduce this behavior. Intriguingly, when $\gamma\ll 1$, we also find flux reversal: cells are swept  {\it against} the moving light field. Simulations and 1D analytics show that such reversal is sensitive to the rate of stochastic reorientation of the swimmers, $k$, and the pattern's duty cycle, $\alpha$. We show that the parameter-dependence of the steady-state downstream accumulation of bacteria can be quite different from the flux, and articulate principles for the `designer transport' of cells using light patterns.  

{\it E. coli} bacteria \cite{schwarz2016escherichia} are $\approx \SI{2}{\micro\meter} \times \SI{1}{\micro\meter}\times \SI{1}{\micro\meter}$ spherocylinders that swim by turning $\approx 7$-$\SI{10}{\micro\meter}$ helical flagella using membrane-embedded rotary motors powered by a PMF of $\approx-\SI{150}{\milli\volt}$ generated by pumping \ce{H+} out of the cells. Unusually, {\it E. coli} can generate a PMF without external nutrients \cite{AdlerEnviron} by using internal resources and \ce{O2}. Without \ce{O2}, swimming ceases \cite{schwarz2016escherichia} unless there is another source of PMF, as in cells expressing PR \cite{Walter2007}, a photon-driven proton pump. Thus, under anaerobic conditions, PR-expressing {\it E. coli} cells swim only when illuminated, which is analogous to synthetic light-activated swimmers \cite{buttinoni2012active,Palacci2013}.

We inserted a PR-bearing plasmid into {\it E. coli} AB1157, and deleted the {\it che}Y gene and the {\it unc} operon encoding the ATP synthase complex to give strain AD10 \cite{Jochen2018}. The former deletion turns wild-type run-and-tumblers into smooth swimmers, while the latter gives fast stopping whenever illumination ceases \cite{Jochen2018}.

Cells suspended in phosphate motility buffer (MB) were diluted to optical density $\approx 6$ at 600 nm ($\approx 0.8$~vol.\% cell bodies~\cite{schwarz2016escherichia}). \SI{2}{\micro\liter} aliquots sealed into \SI{20}{\micro\meter}-thick, $\approx \SI{10}{\milli\meter}$ wide, flat glass capillaries were observed in phase contrast under red illumination using a PF $10\times$/0.3 NA objective on a Nikon TE2000 microscope.
Swimming stopped a few minutes after sealing due to \ce{O2} depletion. 
After leaving these cells in the dark for a further \SI{10}{\minute}, uniform green illumination was turned on (510-\SI{560}nm, corresponding to peak PR absorption; $\approx 5$~\si{\milli\watt\per\square\centi\meter} at the sample). Differential dynamic microscopy (DDM) \cite{WilsonDDM,MartinezDDM} returned an increasing mean swimming speed $\bar{v}$, saturating at $\approx \SI{6.5}{\micro\meter\per\second}$, with standard deviation $\approx \SI{2.5}{\micro\meter\per\second}$ and a fraction $\beta \approx 25\%$ of non-motile bacteria ($\bar{v}$ is averaged over the motile bacteria only). The non-motile bacteria have diffusivity $D_{\rm T} \approx \SI{0.15}{\square\micro\meter\per\second}$. 

A spatial light modulator projected a $4\times 4$ array of static and dynamic patterns onto this initially-uniform field of swimmers. Each featured a central dark square (side $l = \SI{85}{\micro\meter}$) inside an outer square (side $L = \SI{300}{\micro\meter}$), Fig.~\ref{fig:Fig1}d. In the static pattern, the square annulus was uniformly illuminated. The dynamic pattern comprised concentric square annuli of equal width $\Lambda/2= \SI{40}{\micro\meter}$ ($\alpha=0.5$) propagating inwards at speed $u$. The area outside the patterns was dark in all cases. 

\begin{figure}[t]
\includegraphics[width=0.5\textwidth,clip]{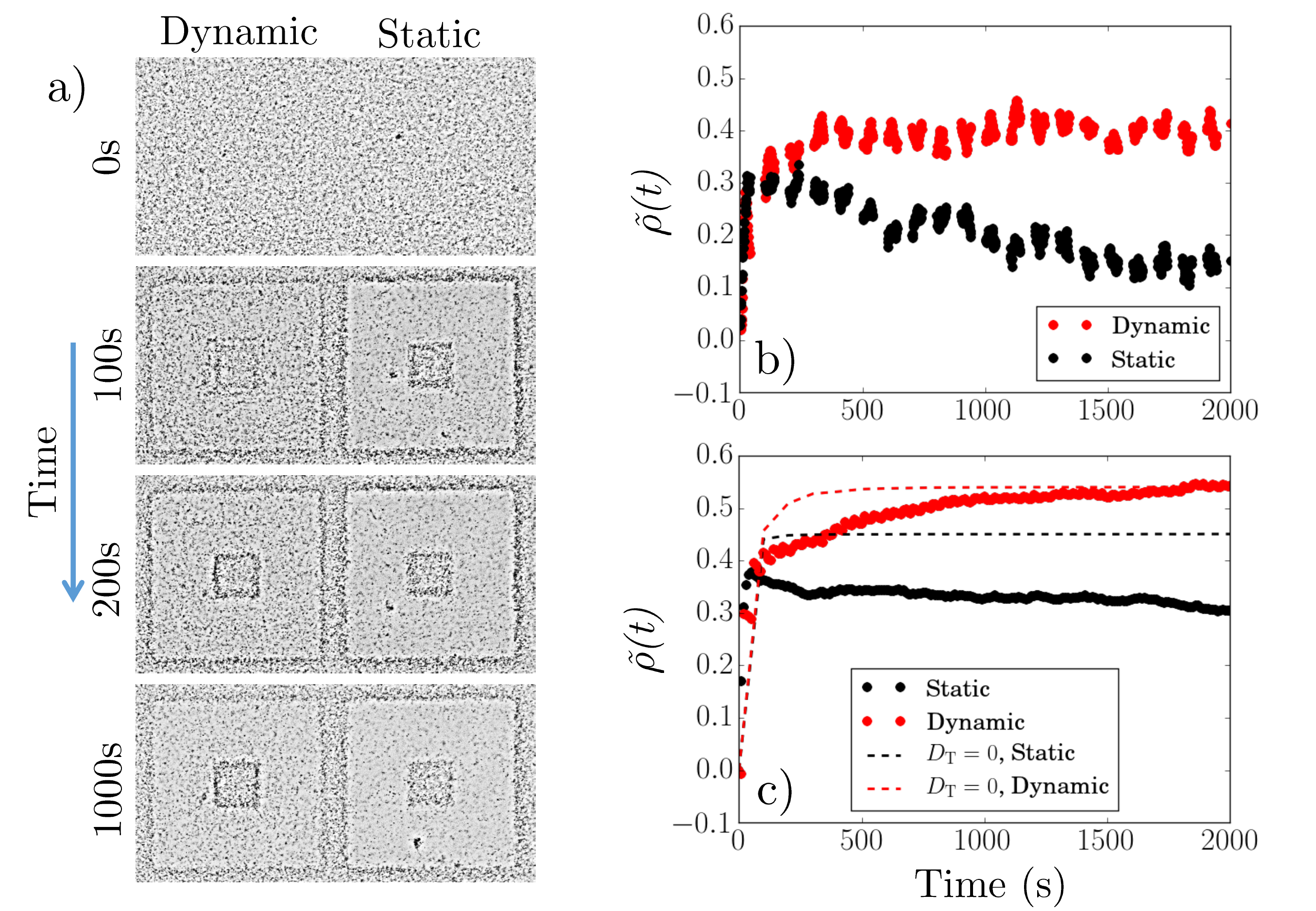}
\caption{\label{fig:Fig2} 
(a) Phase contrast microscopy snapshots of a static pattern, and a dynamic pattern with $u=2.25~\upmu$m/s ($\gamma=u/\bar{v}=0.35$, with $\bar{v}=6.5~\upmu$m/s). (b) Experimental density $\tilde\rho$ at the center of the pattern as a function of time for static and dynamic patterns. (c) 2D simulated $\tilde\rho$ for a dynamic ($u=2.0~\upmu$m/s, $\gamma=0.31$) and a static pattern, with (dots) and without (dashes) thermal motion.}
\end{figure}

In our setup, the local intensity variance, $\sigma^2$, has previously been found to be proportional to the cell density \cite{Jochen2018}. We use this to obtain the relative accumulation of cells which are able to swim (motile) in the inner square:
\begin{equation}
\tilde{\rho}(t)\equiv\frac{\rho_{\rm in}(t)- \rho_{\rm in}(0)}{\rho_{\rm in}(t)+\rho_{\rm in}(0)}=\frac{\sigma^2(t)-\sigma^2(0)}{\sigma^2(t)+\sigma^2(0)(1-2\beta(0))}\,,
\end{equation}
where $\rho_{\rm in}(t)$ is the time-dependent motile cell density in the inner square, and where in the final equality we have assumed zero net transport of non-motile cells
i.e.,~$\beta(t)\sigma^2(t)=\beta(0)\sigma^2(0)$ and $\rho_{\rm in}(t)=\sigma^2(t)-\sigma^2(0)\beta(0)$. Initially,~$\rho_{\rm out}(0) =\rho_{\rm in}(0)= \rho_0$, the uniform motile density everywhere. Subsequently, the outer motile cell density stays approximately constant, i.e., $\rho_{\rm out}(t)\sim\rho_0$, so that $\tilde{\rho}(t)$ also corresponds to the contrast between the inside and outside of the pattern, i.e., $\tilde{\rho}(t)=(\rho_{\rm in}(t)-\rho_{\rm out}(t))/(\rho_{\rm in}(t)+\rho_{\rm out}(t))$. Note that $\tilde\rho = \pm1$ for complete rectification in the inwards or outwards directions respectively. 

Fig.~\ref{fig:Fig2} shows data for $u = \SI{2.25}{\micro\meter\per\second}$ or $\gamma= 0.35$. Imposing the static pattern `paints' a central dark square and the outer square acquires a dark edge, Fig.~\ref{fig:Fig2}a, as previously \cite{Jochen2018}, because cells in the bright (B) annulus swim into the dark (D) central square and outer regions in roughly equal numbers and stop upon arrival. This initial influx of swimmers from the bright annulus gives a sharp increase in $\rho_{\rm in}$ and therefore $\tilde\rho$, Fig.~\ref{fig:Fig2}b ($\bullet$). Thereafter, $\tilde\rho$ decreases slowly as accumulated non-swimmers diffuse back into the bright annulus. Presumably, at long times, a steady state obtains where the B $\to$ D active flux balances the D $\to$ B diffusive flux, so that eventually $\rho_{\rm in}(t \to \infty) \approx \rho_0$ and $\rho_{\rm ann} \ll \rho_0$, while $\rho_{\rm out} = \rho_0$ (outer regions $\approx$ quasi-infinite reservoir).

\begin{figure}[t]
\includegraphics[width=0.5\textwidth,clip]{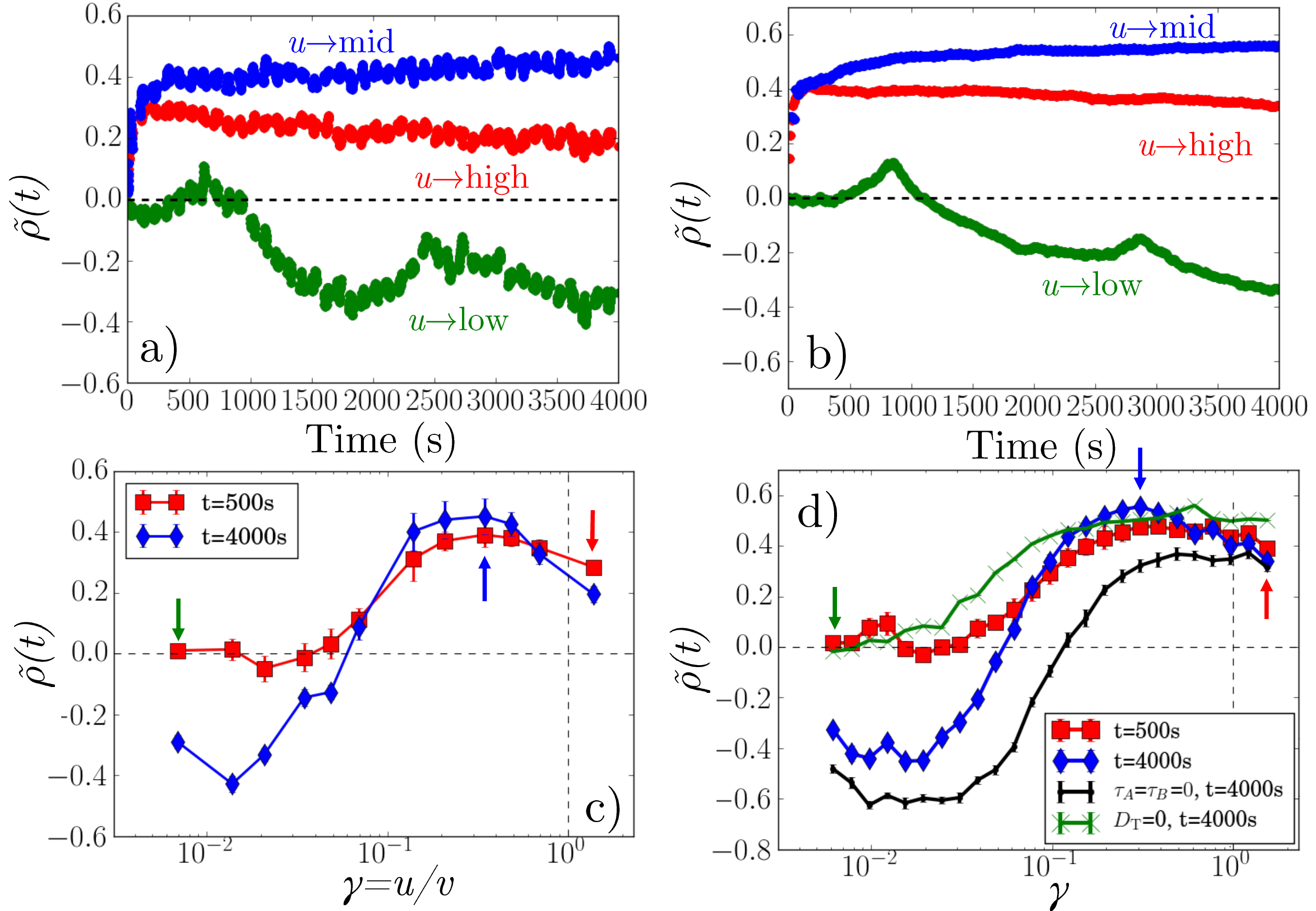}
\caption{\label{fig:Fig3} (a) Experimental time-dependent normalized density $\tilde\rho(t)$ for pattern speeds $u=$0.045, 2.25, 9 $\upmu$m/s ($\gamma=0.007,~0.35,~1.4$). (b) 2D simulations ($\gamma=0.006,~0.31,~1.54$). (c) Experimental $\tilde\rho$ as a function of pattern speed for indicated times from experiments. (d) As in (c) but from simulations, and also results for $D_\mathrm{T}=0$ and $\tau_\mathrm{A}$=$\tau_\mathrm{B}$=0. Arrows in (c) and (d) correspond to the $\gamma$ from (a) and (b).}
\end{figure}

The dynamic light pattern also accumulates cells in the central square, Fig.~\ref{fig:Fig2}a. The difference from the static case
is that $\tilde\rho(t)$ does {\it not} decay after the initial increase, but reaches instead a finite steady-state value. This
accumulation depends non-monotonically on $\gamma = u/v$, Fig.~\ref{fig:Fig3}a, and, counterintuitively, at $\gamma \ll 1$, the accumulation reverses sign: cells are swept {\it out} of the central region ($\tilde\rho < 0$). Both of these features are clearly illustrated in Fig.~\ref{fig:Fig3}c, showing that $\tilde\rho$ as a function of $\gamma$ is peaked at $t =$ \SI{500}{\second} and evolves at \SI{4000}{\second} to $\tilde\rho < 0$ when $\gamma \lesssim 0.05$. 

To explore these features, we simulated $2\times10^4$ non-interacting swimmers in a $600\times600~\upmu$m$^2$ periodic box. Particle speeds $v_i$ are taken from a Schultz distribution (mean = \SI{6.5}{\micro\meter\per\second}, 40\% standard deviation). The dynamics of  particle $i$ obeys
\begin{align}
\dot{\mathbf{r}_i} &=v_{i} A_i B_i\mathbf{p}_i + \sqrt{2D_{\mathrm T}}\boldsymbol{\xi}_{\mathrm T}, \label{eq:rdot}\\
\dot{\theta}_i&=\sqrt{2D_{\mathrm R}} \xi_{\mathrm R}, \label{eq:theta}\\
\dot{A_i}&= (I(\mathbf{r}_i) - A_i ) /\tau_\mathrm{A},\\
\dot{B_i}&= (I(\mathbf{r}_i) - B_i ) /\tau_\mathrm{B},
\end{align}
with $\mathbf{r}_i$ and $\mathbf{p}_i=(\cos\theta_i,\sin\theta_i)$ its position and propulsion direction respectively, and $\boldsymbol{\xi}_{\mathrm T}$, $\xi_{\mathrm R}$ unit-variance Gaussian noise terms (in each direction for $\boldsymbol{\xi}_{\mathrm T}$). Translational, $D_{\mathrm T}=\SI{0.15}{\square\micro\meter\per\second}$, and rotational, $D_{\mathrm R}=\SI{0.05}{\per\second}$, diffusivities reflect experimental values. The light intensity, $I(\mathbf{r}_i,t) =  1$ for bright and 0 for dark. The two dynamical variables $A$ and $B$ reflect the observation\cite{Jochen2018} that two independent processes control the response of our cells to changes in the intensity of external illumination. We measured $\tau_\mathrm{A} = \SI{1.6}{\second}$ and  $\tau_\mathrm{B} = \SI{100}{\second}$. 

Our simulations reproduce the observed rise and decay in $\tilde\rho(t)$ for the static pattern, Fig.~\ref{fig:Fig2}c ($\bullet$). The saturation behavior for the dynamic pattern is also reproduced, Fig.~\ref{fig:Fig2}c ({\color{red} $\bullet$}). In the static case, the simulated $\tilde{\rho}$ decays slower than experiments, by a factor of $\approx 2$, possibly due to our neglect of cell-cell interactions and resulting diffusivity increase due to activity~\cite{jepson2013}; but even so, the semi-quantitative agreement is gratifying given the simplicity of our model. 

Simulations confirm that the diffusion of cells out of the centre was responsible for the decay of $\tilde\rho(t)$ for the static pattern, Fig.~\ref{fig:Fig2}b ($\bullet$).
Setting $D_{\rm T} = 0$ removes the drop in $\tilde\rho(t)$, Fig.~\ref{fig:Fig2}c ($\mbox{-~-~-}$). The effect of removing diffusion on the dynamic pattern is to render the rise to saturation significantly more rapid, Fig.~\ref{fig:Fig2}c ({\color{red} $\mbox{- - -}$}). Thus, a dynamic pattern has to work against thermal diffusion. 

The simulations account well for the dynamics of $\tilde\rho(t)$ at our selected experimental $u$ values, Fig.~\ref{fig:Fig3}b, including the oscillations at low $u$ with characteristic time $\sim \Lambda/ u \sim 2\times\SI{e3}{\second}$. Similarly, $\tilde\rho(\gamma)$ is well reproduced
for both $t =$ \SI{500}{\second} and \SI{4000}{\second}, see Fig.~\ref{fig:Fig3}c-d. In both cases $\gamma \approx 0.3$ is optimal for accumulation, and flux reversal (drainage of the central square) occurs if $\gamma \lesssim 0.1$. Importantly, with $D_{\rm T} = 0$, simulations show no flux reversal (all bacteria transported inwards cannot escape the inner square), while realistic, finite $\tau_{\rm A, B}$ values are needed for quantitative agreement, see Fig.~\ref{fig:Fig3}d, green and black curves, respectively. 

So far we have studied the effect of varying one parameter, the pattern speed $u = \gamma v$ using 2D experiments and simulations. We now construct a 1D analytic model, which allows us to explore the parameter space of the duty cycle, $\alpha$, and the reorientation rate of the bacteria.

We modify a recent theory for light-activated particles in a 1D periodic moving light field~\cite{Leonardo2018} to account for the non-periodic boundaries in our 2D experiments, which permit accumulation. In the original theory, active point particles move right or left at speed $v$ in the light and $v'$ in the dark, but we set $v'=0$ here. Particles reorient independently at rate $k$, which can be viewed as a tumbling or rotational diffusion rate, non-dimensionalised as $\kappa\equiv k \Lambda/u$.  A periodic, square light pattern is imposed, moving at speed $u>0$, in  whose comoving frame are periodic boundary conditions (BCs) at $x=0$ and $x=\Lambda$, with $[0,\alpha \Lambda)$ light and $[\alpha \Lambda, \Lambda)$ dark. 

The Fokker-Planck equation (FPE) for this system with $\alpha = 0.5$ was solved~\cite{Leonardo2018} to yield the average transport velocity $\langle v\rangle$. We extend these results to non-symmetric waveforms ($\alpha\neq 0.5$), and obtain $\langle v_+\rangle$ and $\langle v_-\rangle$, the mean speeds of the $\rightharpoonup$ and $\leftharpoondown$ particle populations, in terms of which $\langle v \rangle = (\langle v_+\rangle - \langle v_-\rangle)/2$. Using these results in a coarse-grained theory, we calculate the concentration difference between either end of a finite square-wave illumination pattern of $n$
periods moving from an `outer' reservoir at $x=0$ to an `inner' reservoir at $x=L=\Lambda n$.

\begin{figure}[t]
\includegraphics[width=0.5\textwidth,clip]{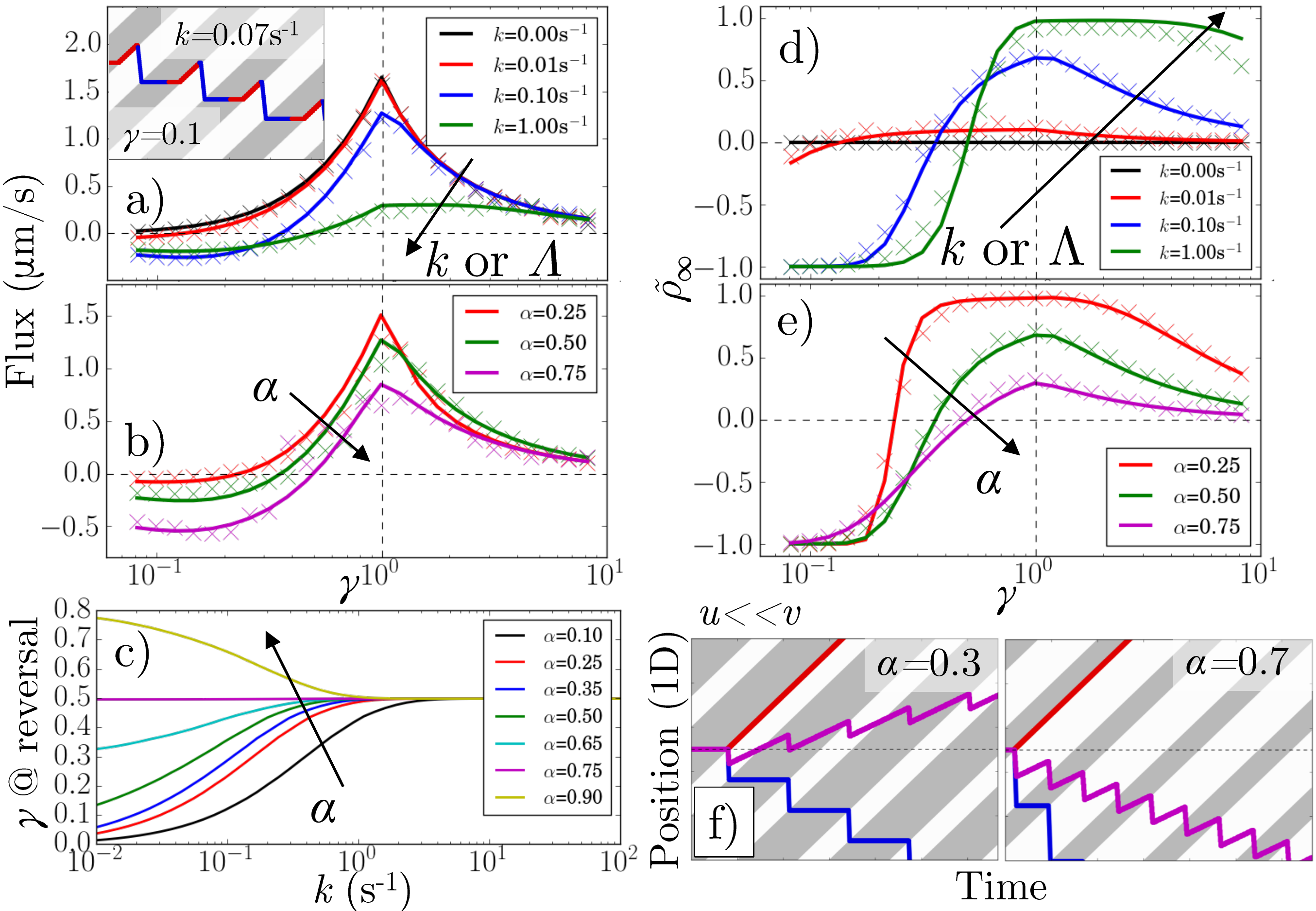}
\caption{\label{fig:Fig4}
(a) 1D theory (lines) and simulations (points) of particle flux {\it vs} pattern speed for $\alpha=0.5$ and $k=0, 0.01, 0.1$ and \SI{1}{\per\second}. Inset: kymograph of a particle tumbling periodically at $k = \SI{0.07}{\per\second}$ with $\alpha=0.5$ and $\gamma=0.1$. (b) As in (a) but for $k= \SI{0.01}{\per\second}$ and $\alpha=0.25, 0.5$ and $0.75$. (c) Predicted critical $\gamma$ for flux reversal, $\gamma^*$, {\it vs} $k$ for different $\alpha$ (in legend). (d) 1D simulation and theory of $\tilde{\rho}$ {\it vs} pattern speed for $\alpha=0.5$ and $k=0, 0.01, 0.1$ and \SI{1}{\per\second}. (e) As in (d) but for $k=\SI{0.01}{\per\second}$ and $\alpha=0.25, 0.5$ and $0.75$. (f) Kymographs for $\alpha=0.3$ and $\alpha=0.7$ with traces of non-tumbling particles moving left (blue) and right (red) for $\gamma\ll 1$. Magenta: average of the traces. }
\end{figure}  

In the moving region, $0<x<L$, we coarse-grain over the periodic dynamics by approximating $\rightharpoonup$ and $\leftharpoondown$ particles as having the respective average speeds, $\langle v_\pm\rangle$, from the periodic BCs model. The resulting FPE for the time-invariant probability densities $\phi_\pm$ in the lab frame is
\begin{align}
0=\frac{\partial \phi_+}{\partial t}=-\langle v_+\rangle \frac{\partial \phi_+}{\partial x}+\frac{k}{2}\left(\phi_--\phi_+\right)\,,\nonumber\\
0=\frac{\partial \phi_+}{\partial t}=\langle v_-\rangle \frac{\partial \phi_-}{\partial x}+\frac{k}{2}\left(\phi_+-\phi_-\right)\,. \label{coarse grain}
\end{align}
BCs account for the reservoirs: the flux $j_+(0)$ of $\rightharpoonup$ particles into the light-pattern at $x=0$ is 
\begin{align}
j_+(0)=\phi_+\langle v_+ \rangle=C\rho_{\rm out}\,,
\end{align}
with $\rho_{\rm out}$ the bacterial density in the $x = 0$ reservoir, and $C$ a constant accounting for the rate at which bacteria exit the reservoirs (the value of $C$ does not affect the steady state provided $C>0$). Similarly, the flux of $\leftharpoonup$ particles out of $x = L$ reservoir is
\begin{align}
j_-(L)=-\phi_-\langle v_- \rangle=-C\rho_{\rm in}\,.
\end{align}
Setting $j_+(0)=-j_-(0)$ and $j_+(L)=-j_-(L)$ maintains constant reservoir densities. Solving Eq.~\eqref{coarse grain} then yields $\tilde{\rho}_\infty=\tilde{\rho}(t\rightarrow\infty)$, the steady-state accumulation parameter 
\begin{align}
\tilde{\rho}_\infty\,=\,\tanh\left(\frac{kn\Lambda\langle v\rangle}{2\langle v_+\rangle\langle v_-\rangle}\right)\,, \label{accumulation}
\end{align}
which is set by a balance between the persistent particle flux, $\langle v\rangle$, and stochastic events where individual particles traverse the pattern against the flux; e.g., when $\langle v_-\rangle\ll\langle v_+\rangle$, $\tilde{\rho}_\infty\sim\tanh\left[kn\Lambda/(4\langle v_-\rangle)\right]$, where $[\ldots]$ is inversely proportional to the probability $\sim\langle v_-\rangle/(kn\Lambda)$ of a particle crossing the entire pattern against the flux without re-orienting. Equivalently, the flux competes with an effective diffusivity $D_{\rm eff}=\langle v_+\rangle\langle v_-\rangle/k$ driving particles down the concentration gradient $(\rho_{\rm in}-\rho_{\rm out})/(n\Lambda)$. 

These theoretical results predict what happens when we vary $\alpha$ and $\kappa$ (by changing $k$ or $\Lambda$ or both), Fig.~\ref{fig:Fig4}. In particular, the flux reverses at a critical $\gamma^* < 1$, Fig.~\ref{fig:Fig4}a,b, that is $k$ and $\alpha$-dependent, Fig.~\ref{fig:Fig4}c. Under our $v' = 0$ conditions, such reversal requires a finite $k$ (whereas for $v'>0$, reversal can occur at $k=0$~\cite{Leonardo2018}). 

The speed reversal at $k>0$ occurs, Fig.~\ref{fig:Fig4}a, because of an asymmetry in the effect of tumbling on the $\rightharpoonup$ and $\leftharpoondown$ parts of the trajectory, which is illustrated in the inset for regular (period $1/k$) tumbling. During $\rightharpoonup$ (red) periods particles are retarded because they become un-trapped from the moving interface and so spend more time static in the dark. The effect on the $\leftharpoondown$ (blue) part of the trajectory is weaker because tumbling does not hinder the rapid runs through the light, until the tumbling rate 
$k\gtrsim (u+v)/(\Lambda\alpha)$.

How the duty cycle $\alpha$ affects reversal, Fig.~\ref{fig:Fig4}b, is illustrated by the kymographs in Fig.~\ref{fig:Fig4}f. For increasing $\alpha$, $\leftharpoonup$ particles spend longer in the light, whereas $\rightharpoonup$ particles trapped at the boundary are unaffected; hence $\gamma^*$ increases with $\alpha$. 

Perhaps counter-intuitively, the accumulation $\tilde\rho_\infty$, is not trivially related to the flux, Figs.~\ref{fig:Fig4}d, e. The case of $\alpha=0.5$ is striking: $\tilde\rho_\infty$ generally increases with increasing $k$ even though the flux decreases. This happens mainly because the probability of a bacterium moving from one resevoir to the other, without tumbling, increases as $k$ decreases. Hence it becomes more difficult to maintain a concentration gradient, and this effect is stronger than the simultaneous decrease of the flux.

To validate our analytics, we simulated a 1D system obeying the same dynamical equations already used for 2D, but with the reorientation, Eq.~\ref{eq:theta}, replaced by a Poissonian tumbling process and $\tau_\mathrm{A}$=$\tau_\mathrm{B}$=0. This reproduced the predicted fluxes exactly, Figs.~\ref{fig:Fig4}a,b, as before~\cite{Leonardo2018}. To extract $\tilde{\rho}$, we used simulations of a finite system, adding a small $D_{\rm T} = 0.01 \upmu {\rm m^2\,s^{-1}}$ to allow the particles to escape $\SI{20}{\micro\meter}$ reservoirs. These simulations only approximately reproduced the theory, Figs.~\ref{fig:Fig4}d,e, which is expected, as the theory
should be exact only in the limit of (i) $n\gg 1$, where the BCs become less important, and (ii) for limited accumulation each cycle, i.e., $k\Lambda\langle v\rangle/(\langle v_+\rangle\langle v_-\rangle)\ll 1$, where it is valid to approximate the system as periodic.

These results suggest that by tuning $\alpha$ and $\kappa$, it should be possible to
control how the accumulation and flux vary with $\gamma$. This is significant because polydispersity in e.g., speed and rotational diffusion rates are unavoidable
and the ideal pattern design will depend on the goal. For example, concentrating all bacteria in a single target region requires bacteria with a wide range of speeds to be strongly rectified in the same direction, corresponding to
low $\alpha$ or high $\Lambda$ in Fig.~\ref{fig:Fig4}d-e. Alternatively, to separate a sample on the basis of speed when there is also polydispersity in tumbling or rotational diffusion, the value of $\alpha=0.75$ would be ideal, as the reversal point is then independent of $k$, see Fig.~\ref{fig:Fig4}c. 


Our results suggest a range of design principles for controlling the flux and accumulation of light-controlled bacteria, and more generally, of active particles in spatiotemporally varying fields, e.g., electrical~\cite{Gangwal2008} or ultrasound~\cite{Rao2015}. These principles will be invaluable in developing applications such as the separation of polydisperse mixtures, the self-assembly of active particles, and the dynamic actuation and control of microscopic machines.

\noindent{\bf{Acknowledgements:}} We thank Teun Vissers and Mike Cates for discussions. NK was part-funded by the EU (H2020-MSCA-IF-2014, ActiDoC No.~654688). AB received funding from UK EPSRC (EP/S001255/1). All except SEG received funding from UK EPSRC (EP/J007404/1) and ERC (Advanced Grant ERC-2013-AdG 340877-PHYSAP). SEG held an EPSRC studentship. The data of the paper will be available on the University of Ediburgh DataShare open access data repository.

\end{document}